\documentclass[aps,prb,twocolumn,longbibliography] {revtex4-2}
\usepackage{amsmath,graphicx,amsfonts}
\usepackage{times}
\usepackage[colorlinks=true,linkcolor=blue,urlcolor=blue,citecolor=blue, breaklinks=true,hypertexnames=false]{hyperref}

\usepackage{amssymb}
\usepackage{bbm}

\usepackage[utf8]{inputenc}
\DeclareUnicodeCharacter{03BC}{\mbox{$\mu$}}

\DeclareUnicodeCharacter{03B4}{$\delta$}
\DeclareUnicodeCharacter{2009}{-}

\DeclareMathOperator{\Ai}{Ai}
\DeclareMathOperator{\Bi}{Bi}

\usepackage[english]{babel}
\addto\captionsenglish{}
\addto\captionsenglish{}

\makeatletter
\def\bbl@set@language#1{%
	\edef\languagename{%
		\ifnum\escapechar=\expandafter`\string#1\@empty
		\else\string#1\@empty\fi}%
	\@ifundefined{babel@language@alias@\languagename}{}{%
		\edef\languagename{\@nameuse{babel@language@alias@\languagename}}%
	}%
	\select@language{\languagename}%
	\expandafter\ifx\csname date\languagename\endcsname\relax\else
	\if@filesw
	\protected@write\@auxout{}{\string\select@language{\languagename}}%
	\bbl@for\bbl@tempa\BabelContentsFiles{%
		\addtocontents{\bbl@tempa}{\xstring\select@language{\languagename}}}%
	\bbl@usehooks{write}{}%
	\fi
	\fi}
\newcommand{\DeclareLanguageAlias}[2]{%
	\global\@namedef{babel@language@alias@#1}{#2}%
}
\makeatother

\DeclareLanguageAlias{en}{english}

\makeatletter
\def\@bibdataout@aps{%
	\immediate\write\@bibdataout{%
		@CONTROL{%
			apsrev41Control%
			\longbibliography@sw{%
				,author="08",editor="1",pages="1",title="0",year="1"%
			}{%
				,author="08",editor="1",pages="1",title="",year="1"%
			}%
		}%
	}%
	\if@filesw \immediate \write \@auxout {\string \citation {apsrev41Control}}\fi 
}
\makeatother

\def\B {\scriptscriptstyle{{B}}}
\def\wkb {\scriptscriptstyle{{WKB}}}

\date\today

\begin{document}
\title{The exchange coupling of a Wigner dimer}

\author{Daniele Lagasco}
\affiliation{Laboratoire de Physique Th\'{e}orique, Univ Toulouse, CNRS, Toulouse, France}
\author{Zoran Ristivojevic}
\affiliation{Laboratoire de Physique Th\'{e}orique, Univ Toulouse, CNRS, Toulouse, France}

\begin{abstract}
We study the exchange coupling in small Wigner crystals confined to one-dimensional space. In particular we concentrate on the simplest nontrivial case of two electrons in a box potential and calculate analytically the energy splitting between the lowest spatially symmetric and antisymmetric states, which is a relevant energy scale for the magnetic properties of the system. In the approximation of a fixed center of mass coordinate, the splitting  decays exponentially with the square root of the distance between the electrons at the leading order. We show that the subleading exponential correction significantly increases the splitting and thus becomes crucial in order to describe correctly the exact numerical data for system sizes that are not astronomically large. Two methods of calculation of the energy splitting are developed. The first is based on the analysis of the exact solution that can be expressed in terms of the Whittaker functions. It applies at all values of the short-distance cutoff played by the width of one-dimensional wire that regularizes the Coulomb potential. The second method is based on the quasiclassical (or Wentzel--Kramers--Brillouin) approximation, which applies only for sufficiently large values of the cutoff. The two methods give identical result in the overlapping region. As a side result, our study gives the energy splitting in a triangular double well potential of inverse ``M'' shape.     
\end{abstract}
\maketitle

\section{Introduction}\label{sec1}

The ground-state properties of an electron system is a fundamental problem in quantum physics that has started to attract the scientific community a century ago and still continues to be at the heart of modern condensed matter physics. The complexity of the problem arises from the interactions that are difficult to treat in such many-body systems. The importance of its understanding is vast, covering various properties of metals and materials. 

In 1934, Eugene Wigner predicted that the long-range interactions between electrons can have a drastic effect, leading to their crystallization \cite{wigner_interaction_1934}. The resulting paradigmatic state is nowadays known as a Wigner crystal. It occurs at low densities when the interaction energy of electrons dominates over the kinetic one. The corresponding dimensionless parameter is $r_s$ and represents the ratio between the average distance between electrons and the effective Bohr radius. At high densities ($r_s \ll 1$) the Coulomb interaction plays a minor role in two- and three-dimensional cases, and the electron gas is described by Landau's Fermi-liquid theory. On the contrary, in the low-density regime ($r_s \gg 1$) the Coulomb interaction prevails, leading to a crystalline phase. There, the effect of interaction is so strong that the energy is minimized for an arranged periodic structure of electrons. This state is characterized by broken translational symmetry, where the electrons are sharply localized around their equilibrium positions. While the spin of electrons does not directly enter to the Coulomb interaction, it plays the role through the Pauli principle and determines the magnetic properties of Wigner crystals. The resulting magnetic order in two- and three-dimensional crystals is \textit{a priori} not known and depends on $r_s$ \cite{drummond_phase_2009,candido_magnetic_2004}.

A system of electrons confined by a strong transverse potential effectively becomes one-dimensional. Interacting electrons in one dimension show strikingly different behavior from their higher-dimensional counterparts \cite{giamarchi}. Due to the enhanced role of quantum fluctuations, there is no possibility for the long-range charge order and therefore there is no phase transition between the crystalline and liquid phases. Instead, a smooth crossover occurs at a certain value $r^*$ of the dimensionless parameter. Nevertheless, the two regimes of high ($r_s \ll r^*$) and low ($r_s \gg r^*$) densities show clear distinctions. In the former, weakly interacting regime the system with $2N$ electrons confined to a box is characterized by the mean density in the ground state that has $N$ peaks corresponding to $N$ pairs, each containing two electrons with the opposite spin projections. This is reminiscent to the Friedel oscillations of the electron density near an impurity. In the latter, Wigner crystal-like regime the situation is quite different. There the interactions are so strong that the electrons are individually separated as much as possible. The resulting density possesses $2N$ peaks.

The Wigner crystal-like regime is also characterized by a magnetic order. Since in one dimension the system of interacting fermions cannot be spin polarized \cite{lieb_theory_1962}, the system always shows antiferromagnetic properties, which is different from the higher-dimensional cases \cite{drummond_phase_2009,candido_magnetic_2004}. This scenario starts to deviate in the case of a weaker transverse confinement where the electrons can form a quasi-one-dimensional zig-zag structure \cite{meyer_wigner_2008}. An electron in this case has several close neighbors and the magnetic phase diagram becomes much richer. In the two-dimensional case, relative importance between various exchange interactions and resulting magnetic properties were studied in several works \cite{roger_multiple_1984,katano_multiple-spin_2000,bernu_exchange_2001}. This picture have been recently complemented by studying the role of interstitial defects and vacancies whose dynamics mediates magnetism on different energy scales than the exchange energies in pure Wigner crystals \cite{kim_interstitial-induced_2022,kim_dynamical_2024}. 

Experimental observation of Wigner crystals has been elusive for a long time due to diverse reasons. Most metals have moderate $r_s$ values that are too far from the critical one needed to reach the crystallization. Achieving very small densities has proven to be very difficult due the appearance of disorder and invasive measurement techniques that destroy the crystalline order. The earliest indications of Wigner crystallization were obtained in the 1980s using the high magnetic-field transport measurements in semiconducting two-dimensional systems \cite{mendez_high-magnetic-field_1983,andrei_observation_1988} as well as in a sheet of electrons on the surface of liquid helium \cite{grimes_evidence_1979}. Very recently, the direct observation of Wigner crystals has been achieved \cite{shapir_imaging_2019,smolenski_signatures_2021,zhou_bilayer_2021,tsui_direct_2024}. In the one-dimensional case, the charge density in the Wigner crystal-like state has been achieved using the carbon nanotube setup, where the system with few electrons has been studied \cite{shapir_imaging_2019}. The latter experimental achievement where the spatial (charge) ordering of electron was directly seen opens the possibility to study their magnetic ordering. At the same time it calls for theoretical studies of magnetic properties of such small systems.

In this paper we study the exchange coupling in small Wigner crystals, which is the most basic parameter that controls the magnetic properties of such systems. Motivated by the experimental achievements \cite{shapir_imaging_2019} where a detection resolution reached the level of a single electron in isolated one-dimensional wells, we consider the simplest nontrivial case with two electrons in a one-dimensional box potential. In this situation the wave function vanishes when any of particles' coordinates coincides with the system ends. The analytical solution of the Schr\"{o}dinger equation for two particles with Coulomb repulsion in such case is still a highly nontrivial problem. Due to the absence of translational symmetry, both coordinates, of the center of mass and the relative one, nontrivially participate in the solution. In order to make analytical progress we consider the approximation of a frozen center of mass, where only the distance between the electrons is the dynamical variable \cite{hausler_correlations_1995,matveev_conductance_2004} that determines the energy levels. The original problem then reduces to the problem of a single quantum particle in a symmetric double well potential with hard walls. 

Physics of a quantum particle in a double-well potential is well known. If the barrier between the wells is high with respect to the zero-point energy in each of the wells, in the first approximation we can consider each of the wells as an isolated system. Then the particle can be localized in either well in the state that corresponds to the one where the other well did not exist. Due to the presence of two identical wells, the particle can be either in one of them or in the other and thus there is a double degeneracy of the ground-state energy. Instead of dealing with the two isolated wave functions centered in each of the wells, we can form their symmetric and antisymmetric combinations. Since the two wells are not completely isolated, there is a small probability for particle tunneling between the wells, which lifts the energy degeneracy. The symmetric state then becomes of a lower energy than the antisymmetric one. The resulting energy splitting typically decays in an exponential way with the distance between the wells. In this paper we study the splitting for the case of a hard-wall double-well potential, which is  less studied than the ``soft-wall'' potential. The important difference between the two is zero probability to find the particle outside the well with the hard-wall condition, in contrast to the other case where the particle has nonzero probability to be anywhere in space. 

The exchange coupling in longer Wigner crystals was studied previously in several works \cite{klironomos_exchange_2005,fogler_exchange_2005,fogler_spin_2005,matveev_conductance_2004,fogler_short-range_2005}. For three and more particles on a ring, the exchange coupling was shown to decay exponentially as $\exp(-\eta\sqrt{r_s})$, where $\eta\approx 2.798$ in the limit of a large number of electrons. The parameter $r_s$ we conveniently define as the ratio of the average distance between the particles and the Bohr radius \footnote{Note that this is different from the conventional definition of the so-called Wigner--Seitz radius that in $d$ dimensions is given by $r_s={(a_{\B})^{-1} (V_d n)^{-1/d}}$. Here $n$ is the electron density, $a_{\B}$ is the Bohr radius, and $V_d$ is the volume of a $d$-dimensional sphere of unit radius. In $d=1$, we have $V_1=2$ and thus $r_s=(2n a_{\B})^{-1}$\cite{fogler_exchange_2005,fogler_spin_2005} }. The variation of $\eta$ with the number of electrons is almost negligible as $\eta$ grows with decreasing the number of electrons toward $\eta\approx 2.801$ for three electrons \cite{fogler_exchange_2005}. On the other hand, for an array of electrons where the positions of all but two exchanging electrons are fixed at their classical equilibrium positions and the two remaining ones have a fixed center of mass coordinate and the relative coordinate as a dynamical variable, it was found $\eta\approx 2.817$ \cite{matveev_conductance_2004}. 

The results of Refs.~\cite{klironomos_exchange_2005,fogler_exchange_2005,fogler_spin_2005,matveev_conductance_2004} however do not apply for small Wigner crystals, or more appropriately molecules, which are well-isolated from the environment. The latter have been realized in the recent experiment \cite{shapir_imaging_2019}. Describing such small Wigner crystals by a one-dimensional system of electrons with Coulomb repulsion subject to zero boundary conditions, here we study their exchange coupling. We show that it decays exponentially at the leading order as $\exp(-\eta\sqrt{r_s})$ with $\eta=\pi$ for a system with two electrons---a Wigner dimer \footnote{Note that for two electrons, their average distance is actually equal to the system size.}. Such decay of the exchange coupling that is faster than the one obtained in systems with a larger number of electrons \cite{klironomos_exchange_2005,fogler_exchange_2005,fogler_spin_2005} is actually slowed down by the subleading exponential term $\exp\left(\tilde \eta\;\! r_s^{1/6}\right)$ with $\tilde\eta=-\pi a_1/2\approx 3.673$, where $a_1$ denotes the first zero of the Airy-$\Ai$ function. The subleading exponential term has a dramatic effect. For example, for $r_s=45$ it gives an increase of the exchange coupling by a factor of $1019$ than one would naively have accounting only for the leading-order term. Moreover for $r_s$ between $5$ and $200$, the product of the two exponents in systems with zero boundary conditions is larger than the exponent for the other setup at least by a factor of $50$.

The rest of the paper is organized as follows. In Sec.~\ref{sec2} we give a mathematical formulation of the problem that reduces to the eigenvalue problem of a single particle that experiences the Coulomb potential regularized by a cutoff to prevent the potential divergence at zero distance. A formal solution of the problem in terms of the Whittaker functions is given in Sec.~\ref{sec3}. The analysis of the formal solution in the case of a small cutoff is given in Sec.~\ref{sec4} that is followed by Sec.~\ref{sec5} where the case of a larger cutoff is considered. The solution of the eigenvalue problem in the quasiclassical or Wentzel--Kramers--Brillouin (WKB) approximation is presented in Sec.~\ref{sec6}, which shows the overlap with the results of Sec.~\ref{sec5}. In Sec.~\ref{sec7}, the discussion of obtained results is given.

\section{Formulation of the problem}\label{sec2}

We study two nonrelativistic electrons confined to one-dimensional system of the size $L$. The Schr\"{o}dinger equation for this problem reduces to the eigenvalue problem
\begin{align}\label{eq:SE}
\left[-\frac{\hbar^2}{2m}\left(\frac{\partial^2}{\partial x_1^2} +\frac{\partial^2}{\partial x_2^2}\right) +V(x_1-x_2)-\mathcal{E}\right]\Psi(x_1,x_2)=0,
\end{align}
where $m$ is the mass of electrons, $V(x_1-x_2)$ accounts for their Coulomb repulsion, and $x_1$ and $x_2$ are the electron positions. We consider the system with hard-wall boundary conditions. This means, the wave function $\Psi(x_1,x_2)$ vanishes if one of the coordinates $x_1$ or $x_2$ coincides with $0$ or $L$, which are taken to be the coordinates of the two boundaries. Our goal is to calculate the energy splitting between the two lowest energy states for the problem (\ref{eq:SE}). The electron spin is here implicitly taken into account as the two lowest energy states will be symmetric and antisymmetric wave functions with respect to the exchange of spatial coordinates. Different symmetries of the spatial part of the wave function will be combined with different spin symmetries of the two lowest states that will make the total wave function of electrons (the one involving spin and spatial degrees of freedom) antisymmetric to the exchange of its coordinates.

Introducing the relative coordinate $x=x_1-x_2$ and the one describing the center of mass, $X=(x_1+x_2)/2$, the eigenvalue problem (\ref{eq:SE}) becomes
\begin{align}
\left[-\frac{\hbar^2}{4m}\frac{\partial^2}{\partial X^2} -\frac{\hbar^2}{m}\frac{\partial^2}{\partial x^2}+V(x)\right] \Psi\left(X+\frac{x}{2},X-\frac{x}{2}\right)\notag\\
=\mathcal{E} \Psi\left(X+\frac{x}{2},X-\frac{x}{2}\right).
\end{align} 
We study the case of a strong potential. In this situation, the system is close to its classical configuration with electrons near $0$ and $L$. 

In the following we neglect the motion along the perpendicular $X$ direction and determine the energy splitting. Note that this is an uncontrolled approximation that should be \textit{a posteriori} examined. In this case $X=L/2$ and the only dynamical variable is the coordinate $x$. This amounts to solving
\begin{subequations}\label{eq:SEeffective}
\begin{align}\label{eq:SEeffectiveonly}
-\frac{\hbar^2}{m}\frac{d^2\psi(x)}{dx^2}+V(x)\psi(x)=E\psi(x)
\end{align}
subject to
\begin{align}\label{eq:SEbc}
\psi(-L)=\psi(L)=0.
\end{align}
\end{subequations}
Equation (\ref{eq:SEeffectiveonly}) with the boundary condition (\ref{eq:SEbc}) describes an effective single-particle problem in the potential $V(x)$, with $-L\le x\le L$. Our goal is to find the splitting between its two lowest energy states. 

Some general properties of the eigenfunctions for the problem (\ref{eq:SEeffective}) follow from general considerations. The potential $V(x)$ is an even function. In this case the eigenfunctions can be classified with respect to the parity. They can be even, in which case $f(x)=f(-x)$, and odd functions, in which case $f(x)=-f(-x)$  is satisfied. For non-pathological forms of $V(x)$, the ground state wave function will have no nodes and thus be even, while the first excited state should have one node and be odd.

The prime candidate to describe Coulomb repulsion between electrons is a pure Coulomb potential $e^2/|x|$, where $e$ is the electron charge. This potential is, however, singular at $x=0$. Moreover, the pure Coulomb potential is a non-integrable function up to the singularity and it thus represents an impenetrable barrier in one dimension \cite{andrews_singular_1976}. The resulting wave functions in such case must vanish at $x=0$. The wave function $\psi_{>}(x)$ that vanishes at $x=0$ and $x=L$ can be extended to negative $-L\le x<0$ as an even function, $\psi_{<}(x)=\psi_{>}(-x)$, or as an odd function, $\psi_{<}(x)=-\psi_{>}(-x)$. Both extensions will have the same energy, resulting with zero splitting. 

The pure Coulomb potential is, however, pathological, which can be seen as the presence of a node in the ground-state wave function. On the other hand, its functional form at very short distances deviates due to other physical effects, which in our case is the transverse dimension of the one-dimensional system, i.e., the width of the wire \cite{klironomos_exchange_2005}. It is therefore more appropriate to adopt another model of Coulomb potential that is given by
\begin{align}\label{eq:Coulombinteraction}
V(x)=\frac{e^2}{w+|x|}.
\end{align}
The model described by the potential (\ref{eq:Coulombinteraction}) is useful as it is analytically tractable yet capturing well the realistic effects. Indeed, at very short distances, $V(x)$ is not divergent as it is cut off by the width of the wire $w$, acquiring a constant value. On the other hand, at long separations $|x|\gg w$, $V(x)$ has a characteristic $1/|x|$ tail. For the potential (\ref{eq:Coulombinteraction}), the eigenvalue problem (\ref{eq:SEeffective}) will have a ground-state wave function $\psi_0(x)$ that is an even function without nodes. Therefore, it will necessarily be characterized by $\psi_0(0)\neq 0$. On the other hand, the first excited state will be described by an odd wave function $\psi_1(x)$ with one node at $x=0$, i.e., $\psi_1(0)=0$. As $w$ tends to zero, $\psi_0(0)$ will approach zero. At the same time, $\psi_0(x)$ will approach $\psi_1(x)$ up to a sign, and the two eigenvalues will become degenerate. 

Instead of seeking for the two lowest energy states $\psi_0(x)$ and $\psi_1(x)$ of the eigenvalue problem (\ref{eq:SEeffectiveonly}) for the potential (\ref{eq:Coulombinteraction}) at $-L\le x\le L$, it suffices to consider positive $0\le x\le L$ and search for the nodeless solutions that obeys
\begin{align}\label{eq:evensolution}
\psi_0'(0)=0,\quad \psi_0(L)=0
\end{align}
for the even solution and
\begin{align}\label{eq:oddsolution}
\psi_1(0)=0,\quad \psi_1(L)=0
\end{align}
for the odd solution. The first boundary condition in Eq.~(\ref{eq:oddsolution}) is obvious from the definition of odd functions. On the other hand, the first boundary condition of Eq.~(\ref{eq:evensolution}) can be obtained as follows. After performing the integration of Eq.~(\ref{eq:SEeffectiveonly}) between $-\delta$ and $\delta$, in the limit $\delta\to 0^+$ we obtain $\psi_0'(0^+)=\psi_0'(0^-)$. In combination with $\psi_0(x)=\psi_0(-x)$, Eq.~(\ref{eq:evensolution}) then follows.

Eventually we can simplify the notation by introducing the dimensionless units. Expressing the lengths in units of the Bohr radius 
\begin{align}
a_{\B}=\frac{\hbar^2}{me^2},
\end{align}
we can introduce the dimensionless parameters
\begin{align}\label{eq:units}
r_s=\frac{L}{a_{\B}},\quad \tilde w=\frac{w}{a_{\B}},\quad \epsilon=4E\frac{ma_{\B}^2}{\hbar^2},
\end{align}
as well as
\begin{align}
y=\frac{x}{a_{\B}},\quad \chi(y)=\psi(x).
\end{align}
The eigenvalue problem then becomes
\begin{align}\label{eq:EVP}
-\frac{d^2\chi(y)}{dy^2}+\frac{1}{\tilde w+y}\chi(y)=\frac{\epsilon}{4}\chi(y).
\end{align}
For the even solution we require
\begin{align}
\chi_0'(0)=0,\quad \chi_0(r_s)=0,
\end{align}
and for the odd one
\begin{align}
\chi_1(0)=0,\quad \chi_1(r_s)=0.
\end{align}
We note that the case of strong potential in fact denotes the case $L\gg a_{\B}$, i.e., $r_s\gg 1$.

\section{Formal solution in terms of the Whittaker functions}\label{sec3}

Upon the shift of coordinates $z=\tilde w+y$, Eq.~(\ref{eq:EVP}) becomes the   second-order differential equation 
\begin{align}\label{eq:WDE}
-\frac{\partial^2 f(z,\epsilon)}{\partial z^2}+\frac{1}{z}f(z,\epsilon)=\frac{\epsilon}{4}f(z,\epsilon)
\end{align}
with $z>0$, which corresponds to a special form of the Whittaker equation \cite{whittaker_course_1996}. It has two independent solutions. They are expressed in terms of two hypergeometric functions known as the Whittaker functions, $M_{\frac{i}{\sqrt\epsilon},\frac{1}{2}}(i \sqrt{\epsilon}\,z)$ and $W_{\frac{i}{\sqrt\epsilon},\frac{1}{2}}(i \sqrt{\epsilon}\,z)$. In our case of $z>0$ and $\epsilon>0$, the first function is purely imaginary and the second one is complex. Instead of dealing with them, we find it convenient to use another set of real independent solutions given by
\begin{subequations}\label{eq:Whitt}
\begin{align}
f_1(z,\epsilon)={}&\frac{1}{2}\left[W_{\frac{i}{\sqrt\epsilon},\frac{1}{2}}(i \sqrt{\epsilon}\,z)+W_{-\frac{i}{\sqrt\epsilon},\frac{1}{2}}(-i \sqrt{\epsilon}\,z)\right],\\
f_2(z,\epsilon)={}&\frac{1}{2i}\left[W_{\frac{i}{\sqrt\epsilon},\frac{1}{2}}(i \sqrt{\epsilon}\,z)-W_{-\frac{i}{\sqrt\epsilon},\frac{1}{2}}(-i \sqrt{\epsilon}\,z)\right].\!
\end{align}
\end{subequations}
We note that $f_1$ and $f_2$ can be expressed as the real and imaginary parts of $W_{\frac{i}{\sqrt\epsilon},\frac{1}{2}}(i \sqrt{\epsilon}\,z)$. Their Wronskian is given by 
\begin{align}\label{eq:Wronskian}
\mathcal{W}_z\{f_1(z,\epsilon),f_2(z,\epsilon)\}={}&\begin{vmatrix}
f_1(z,\epsilon) & f_2(z,\epsilon)\\
\frac{\partial}{\partial z} f_1(z,\epsilon) & \frac{\partial }{\partial z} f_2(z,\epsilon)\\
\end{vmatrix}\notag\\
={}&-\frac{\sqrt\epsilon}{2}\exp\left(-\frac{\pi}{\sqrt\epsilon}\right).
\end{align}
We note that for $z<0$, Eqs.~(\ref{eq:Whitt}) are also the solutions of Eq.~(\ref{eq:WDE}). However, in this case the Wronskian is $-\frac{\sqrt\epsilon}{2}e^{\pi/\sqrt\epsilon}$.

We are now in a position to state the solutions of Eq.~(\ref{eq:EVP}) that are given as linear combinations of the solutions (\ref{eq:Whitt}). Accounting for the boundary condition at $y=0$, the even one is given by
\begin{align}
\chi_0(y)=f_1(y+\tilde w,\epsilon_0)-\frac{\frac{\partial}{\partial\tilde w}f_1(\tilde w,\epsilon_0)}{\frac{\partial}{\partial\tilde w}f_2(\tilde w,\epsilon_0)}f_2(y+\tilde w,\epsilon_0),
\end{align}
where the corresponding energy $\epsilon_0$ should be obtained from the boundary condition $\chi_0(r_s)=0$, i.e.,
\begin{align}\label{eq:eps0}
\frac{f_1(r_s+\tilde w,\epsilon_0)}{f_2(r_s+\tilde w,\epsilon_0)}=\frac{\frac{\partial}{\partial\tilde w}f_1(\tilde w,\epsilon_0)}{\frac{\partial}{\partial\tilde w} f_2(\tilde w,\epsilon_0)} .
\end{align}
The latter equation has many solutions that correspond to many even excited states of the problem. We will seek the one that has the smallest (positive) energy $\epsilon_0$.
In a similar way we can obtain the odd solution that is given by
\begin{align}
\chi_1(y)=f_1(\tilde w+y,\epsilon_1)-\frac{f_1(\tilde w,\epsilon_1)}{f_2(\tilde w,\epsilon_1)}f_2(\tilde w+y,\epsilon_1).
\end{align}
Here the energy should be found from 
\begin{align}\label{eq:eps1}
\frac{f_1(r_s+\tilde w,\epsilon_1)}{f_2(r_s+\tilde w,\epsilon_1)}=\frac{f_1(\tilde w,\epsilon_1)}{f_2(\tilde w,\epsilon_1)}.
\end{align}
Again, among many solutions we will seek the one with the smallest (positive) energy $\epsilon_1$. Once Eqs.~(\ref{eq:eps0}) and (\ref{eq:eps1}) are solved, the energy splitting for the problem (\ref{eq:SEeffective}) with the potential (\ref{eq:Coulombinteraction}) is given by
\begin{align}\label{eq:Delta}
\Delta=\frac{\hbar^2}{4ma_{\B}^2}(\epsilon_1-\epsilon_0).
\end{align}
Therefore, the energy splitting $\Delta$ has a nontrivial dependence on $r_s$ and $\tilde w$. The exchange coupling $J$ of two electrons in our system is related to the energy splitting $\Delta$. In the ground state the two electrons are described by the symmetric spatial wave function and antisymmetric spin one corresponding to the total spin $0$. In the first excited state, the spatial part of the wave function is antisymmetric and the spin one symmetric with the total spin  $1$. The effective Hamiltonian that describes this is up to a constant given by $J\, \mathbf{S}_0\cdot \mathbf{S}_1$ in terms of spin operators, which gives the identification $J=\Delta$. 

\section{Asymptotic solution for the energy splitting at small $\tilde w$}\label{sec4}

The energy splitting $\Delta$ can be easily calculated numerically from the exact solution (\ref{eq:eps0}) and (\ref{eq:eps1}) at any $r_s$ and $\tilde w$. Obtaining the analytical result is, however, significantly more involved. Let us find the leading-order asymptotics in the regime of strong repulsion, corresponding to $r_s\gg1$. 

Closely looking at Eq.~(\ref{eq:WDE}) we can observe that the function $f(z,\epsilon)$ has a turning point at $z=4/\epsilon$. In this case the second derivative vanishes, $\frac{\partial^2 f(z,\epsilon)}{\partial z^2}=0$. There is qualitatively different behavior of the solutions (\ref{eq:Whitt}) for $0<z<4/\epsilon$ and $z>4/\epsilon$. In the former case they do not have zeros, while in the latter they oscillate around zero. For our interest of finding the two lowest-energy eigenfunctions, we need to study the behavior around the smallest $z$ where they vanish. In order to do it, we will employ the asymptotic expansion of the Whittaker function entering Eqs.~(\ref{eq:Whitt}) at small $\epsilon$ derived by \citet{erdelyi_asymptotic_1957}. For $z>4/\epsilon$ we can express it as a combination of the Airy functions as
\begin{align}\label{eq:erdelyi}
W_{\frac{i}{\sqrt\epsilon},\frac{1}{2}}&(i \sqrt{\epsilon}\,z)\simeq \sqrt\pi \left(\frac{2}{\sqrt\epsilon}\right)^{1/6}
\exp\bigglb(-\frac{\pi}{2\sqrt{\epsilon}}+i \alpha(\epsilon)\biggrb)\notag\\ &\times\frac{1}{\sqrt{\phi'\left(\frac{\epsilon\;\! z}{4}\right)}}\Biglb[\Ai\biglb(g(\epsilon,z) \bigrb)+i \Bi\biglb(g(\epsilon,z) \bigrb)\Bigrb],
\end{align}
with the abbreviations
\begin{gather}\label{eq:g}
	g(\epsilon,z)=-\left(\frac{2}{\sqrt{\epsilon}}\right)^{2/3}\phi\left(\frac{\epsilon\;\! z }{4}\right),\\
	\alpha(\epsilon)=\frac{1}{\sqrt\epsilon}\ln\frac{1}{e\sqrt\epsilon}-\frac{\pi}{4}.
\end{gather}
Here
\begin{align}\label{eq:phi>}
\phi(x)= \left[\frac{3}{2}\sqrt{x(x-1)}-\frac{3}{2}\ln\left(\sqrt{x}+\sqrt{x-1}\right)\right]^{2/3}
\end{align}
for $x\ge 1$. Note that $\phi(x)>0$ for $x>1$. Equation (\ref{eq:erdelyi}) also applies for positive, not very small $z$ that are left from the turning point, $z<4/\epsilon$. However, in this case $\phi(x)$ has a different form,
\begin{align}\label{eq:phi<}
\phi(x)= -\left[\frac{3}{2}\arctan\sqrt{\frac{1-x}{x}}-\frac{3}{2}\sqrt{x(1-x)}\;\right]^{2/3},
\end{align}
where $x<1$.
Note that $\phi(x)<0$ for $x<1$. The function $\phi(x)$ is continuous and smooth at $x=1$. For $x$ around $1$, we note the expansion
\begin{align}\label{eq:phiexp}
\phi(x)=x-1+\mathcal{O}\biglb((x-1)^2\bigrb),
\end{align}
in which case Eq.~(\ref{eq:erdelyi}) simplifies. Its $z$-dependent part can be simply understood as a linear combination of two Airy functions that can be obtained by solving  Eq.~(\ref{eq:WDE}) after the linearization of its second term around the turning point.

The ratio that enters Eqs.~(\ref{eq:eps0}) and (\ref{eq:eps1}) can be evaluated using the obvious property
\begin{align}
\frac{f_1(z,\epsilon)}{f_2(z,\epsilon)}=\frac{\mathrm{Re}\Biglb( W_{\frac{i}{\sqrt\epsilon},\frac{1}{2}}(i \sqrt{\epsilon}\,z)\Bigrb) }{\mathrm{Im}\Biglb( W_{\frac{i}{\sqrt\epsilon},\frac{1}{2}}(i \sqrt{\epsilon}\,z)\Bigrb) }=\tan\left(\frac{\pi}{2}-\varphi\right),
\end{align}
where 
\begin{align}
\varphi=\arg W_{\frac{i}{\sqrt\epsilon},\frac{1}{2}}(i \sqrt{\epsilon}\,z)
\end{align}
denotes the argument of the complex Whittaker function. For the form of Eq.~(\ref{eq:erdelyi}), we obtain
\begin{align}\label{eq:ratioerdelyi}
\frac{f_1(z,\epsilon)}{f_2(z,\epsilon)}=\frac{\Ai\biglb(g(\epsilon,z) \bigrb)-\tan \biglb(\alpha(\epsilon)\bigrb) \Bi\biglb(g(\epsilon,z) \bigrb)} {\tan \biglb(\alpha(\epsilon)\bigrb)\Ai\biglb(g(\epsilon,z) \bigrb)+\Bi\biglb(g(\epsilon,z) \bigrb)}.
\end{align}

Let us now study the right-hand sides of Eqs.~(\ref{eq:eps0}) and (\ref{eq:eps1}). We will do it in the limit of thin wires, $\tilde w\ll 1$. The expansion at small $z$ and then at small $\epsilon$ of $W$-solution leads to
\begin{align}\label{eq:Wexpansion}
&W_{\frac{i}{\sqrt\epsilon},\frac{1}{2}}(i \sqrt{\epsilon}\,z)\simeq\frac{\sqrt{1+\frac{\epsilon}{144}}}{\sqrt{2\pi}} \epsilon^{1/4} [1+z(\ln z -1+2\gamma)]\notag\\
&\times  \exp\Bigglb(\frac{\pi}{2\sqrt{\epsilon}}+i \left[\alpha(\epsilon)+\frac{\pi}{2}-\arctan\left(\frac{\sqrt\epsilon}{12}\right)\;\!\right]\Biggrb),
\end{align}
where $\gamma$ is the Euler constant. The right-hand side of Eq.~(\ref{eq:eps1}) has a finite limit at small $\tilde w$. This is expected, as odd solution vanishes and is not affected by the singularity of the pure Coulomb potential. We thus have
\begin{align}\label{eq:f1of2}
\frac{f_1(\tilde w,\epsilon)}{f_2(\tilde w,\epsilon)}\simeq \tan\Bigglb(\!\arctan\biggl(\!\frac{\sqrt\epsilon}{12}\biggr)-\alpha(\epsilon)\!\Biggrb).
\end{align}
We could also try to evaluate the right-hand side of Eq.~(\ref{eq:eps0}) using Eq.~(\ref{eq:Wexpansion}). The result would be identical to the right-hand side of Eq.~(\ref{eq:f1of2}). This means $\epsilon_0=\epsilon_1$ and thus there is no energy splitting, $\Delta=0$. The conundrum is that the expression for $\Delta$ is nonperturbative in energy (or $r_s$) and thus one has to find a way to treat such smallness. Hopefully this can be achieved using the Wronskian (\ref{eq:Wronskian}). It leads to the exact relation
\begin{align}\label{eq:exactfromW}
\frac{\frac{\partial}{\partial \tilde w}f_1(\tilde w,\epsilon)}{\frac{\partial}{\partial \tilde w} f_2(\tilde w,\epsilon)}=\frac{f_1(\tilde w,\epsilon)}{f_2(\tilde w,\epsilon)} +\frac{\sqrt\epsilon\;\!\exp\left(-\frac{\pi}{\sqrt\epsilon}\right)}{2f_2(\tilde{w},\epsilon) \frac{\partial}{\partial \tilde w} f_2(\tilde w,\epsilon) }\;\!.
\end{align}
Equation (\ref{eq:exactfromW}) in combination with the asymptotic expression (\ref{eq:f1of2}) and the expansion (\ref{eq:Wexpansion}) enables us to find
\begin{gather}\label{eq:df1odf2}
\frac{\frac{\partial}{\partial \tilde w}f_1(\tilde w,\epsilon)}{\frac{\partial}{\partial \tilde w} f_2(\tilde w,\epsilon)}\simeq \tan\Bigglb(\!\arctan\biggl(\!\frac{\sqrt\epsilon}{12}\biggr)-\alpha(\epsilon)-F(\epsilon,\tilde w)\!\Biggrb)
\end{gather}
with
\begin{gather}\label{eq:F}
 F(\epsilon,\tilde{w})=\frac{\pi\;\!\exp\left(-\frac{2\pi}{\sqrt\epsilon}\right)}{\ln\frac{1}{\tilde we^{2\gamma}}}.
\end{gather}
Here we account for the leading-order with respect to the exponentially small term contained in $F$.

We can now solve Eqs.~(\ref{eq:eps0}) and (\ref{eq:eps1}) at $r_s\gg 1$. For the latter we use Eqs.~(\ref{eq:f1of2}) and (\ref{eq:ratioerdelyi}), yielding
\begin{align}\label{eq:eps1final}
\Ai\biglb(g(\epsilon_1,r_s) \bigrb)=\frac{\sqrt{\epsilon_1}}{12}\Bi\biglb(g(\epsilon_1,r_s)\bigrb).
\end{align}
Here we naturally assumed $r_s\gg\tilde w$. The lowest-energy solution of Eq.~(\ref{eq:eps1final}) is given by
\begin{align}\label{eq:eps1solution}
\epsilon_1=\frac{4}{r_s}-\frac{4a_1}{r_s^{4/3}}+\mathcal{O}(r_s^{-5/3}).
\end{align}
Here $a_1\approx-2.338$ denotes the first zero of the Airy-$\Ai$ function, i.e., $\Ai(a_1)=0$. Note that for the first two terms of the expansion in powers of $r_s\gg 1$ in the result (\ref{eq:eps1solution}), the right-hand side of Eq.~(\ref{eq:eps1final}) can be set to $0$. Moreover, accounting for the first neglected term in the expansion (\ref{eq:erdelyi}) would produce additional terms on the right-hand side of Eq.~(\ref{eq:eps1final}) that are proportional to $\sqrt{\epsilon_1}$. The origin of the right-hand side of Eq.~(\ref{eq:eps1final}) is the arctangent of Eqs.~(\ref{eq:Wexpansion}) that we kept, but for the purpose of obtaining our results (\ref{eq:eps1solution}) and (\ref{eq:Deltafinal}), it can be safely replaced by $0$.

Let us now consider Eq.~(\ref{eq:eps0}). Using analogous manipulations as for Eq.~(\ref{eq:eps1final}), the expression (\ref{eq:eps0}) simplifies to
\begin{align}\label{eq:eps0final}
\Ai\biglb(g(\epsilon_0,r_s) \bigrb)=\left[\frac{\sqrt{\epsilon_0}}{12}-F(\epsilon_0,\tilde w)\right]\Bi\biglb(g(\epsilon_0,r_s)\bigrb).
\end{align}
Due to the exponentially small contribution arising from $F$, see Eq.~(\ref{eq:F}), it is difficult to solve analytically Eq.~(\ref{eq:eps0final}) alone. However, the difference between its solution $\epsilon_0$ and the solution $\epsilon_1$ of Eq.~(\ref{eq:eps1final}) can be sought perturbatively in $F$, resulting with
\begin{align}
\epsilon_1-\epsilon_0\simeq \frac{-\Bi(a_1)}{2^{2/3}\Ai'(a_1)}\frac{\epsilon_1^{4/3} F(\epsilon_1,\tilde{w})}{\frac{r_s\epsilon_1}{4} \phi'\left(\frac{ r_s \epsilon_1}{4}\right)-\frac{1}{3} \phi\left(\frac{ r_s \epsilon_1}{4}\right)}.
\end{align}
The series expansion (\ref{eq:eps1solution}) enables us to account for the two leading-order terms in the exponent and for the leading-order term in the pre-exponential factor at large $r_s$. The energy splitting then becomes
\begin{align}\label{eq:Deltafinal}
\epsilon_1-\epsilon_0={}&\frac{-4\pi\Bi(a_1)}{\Ai'(a_1)} \frac{\exp\left(-\pi\sqrt{r_s}-\frac{\pi a_1}{2}r_s^{1/6}\right)}{r_s^{4/3} \ln\frac{1}{\tilde{w}e^{2\gamma}}},
\end{align}
where we note the numerical value of the prefactor $\frac{-4\pi\Bi(a_1)}{\Ai'(a_1)}\approx 8.135$.

\begin{figure}
\includegraphics[width=0.95\columnwidth]{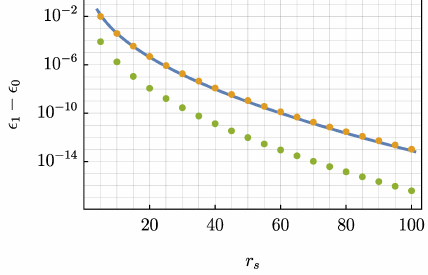}
\caption{Plot of the energy gap as a function of $r_s$ for $\tilde{w}=10^{-5}$. The curve represents the exact value obtained by numerically solving Eqs.~(\ref{eq:eps0}) and (\ref{eq:eps1}). The dots that are on the curve are obtained from the analytical expression (\ref{eq:Deltafinal}). The dots that are outside the curve are obtained from Eq.~(\ref{eq:Deltafinal}) where in the exponential function only the leading term is kept. The latter dotted curve illustrates the important role of the correction term in the exponent.}\label{fig1}
\end{figure}

Equation (\ref{eq:Deltafinal}) gives the analytical expression for the energy splitting at $r_s\gg 1$ and $\tilde w\ll 1$. Several comments are in order. Equation (\ref{eq:Deltafinal}) shows that the energy splitting $\Delta\sim \epsilon_1-\epsilon_0$ between the lowest eigenvalues corresponding to even and odd eigenfunctions vanishes if $\tilde{w}=0$. This is expected as the Coulomb potential in this case is not cut off at shortest distances and in some sense becomes pathological. At $\tilde w=0$, the Coulomb potential is impenetrable, which imposes the nullification of all the eigenfunctions at the origin and thus $\Delta=0$. At $\tilde w>0$, the dominant behavior of $\Delta$ is the exponential decay that occurs as $r_s$ is increased. Note that we have found the correction to the leading term in the exponent, which is very important for practical purposes. The leading term grows as $\sqrt{r_s}$, while the correction also shows a growth as $r_s^{1/6}$, although a slower one. Further subleading terms in the exponent are not crucial as they scale with negative powers of $r_s$ and thus tend to zero. They nevertheless correct the pre-exponential factor $r_s^{-4/3}$, which we found at the leading order. In Fig.~\ref{fig1} we show the comparison between the exact result for the energy splitting (or gap) and the derived analytical expression (\ref{eq:Deltafinal}).

\section{Energy splitting at any $\tilde w$}\label{sec5}

The energy splitting can also be calculated without relying on the smallness of $\tilde w$. Let us consider the exact solution (\ref{eq:eps0}) and (\ref{eq:eps1}) at any $r_s$ and $\tilde w$, which determine the two lowest energies $\epsilon_0$ and $\epsilon_1$. Accounting for the exact relation (\ref{eq:exactfromW}) that involves the Wronskian, Eq.~(\ref{eq:eps0}) becomes
\begin{align}\label{eq:eps0wronskian}
	\frac{f_1(r_s+\tilde w,\epsilon_0)}{f_2(r_s+\tilde w,\epsilon_0)}=\frac{f_1(\tilde w,\epsilon_0)}{f_2(\tilde w,\epsilon_0)} +\frac{\sqrt{\epsilon_0}\;\!\exp\left(-\frac{\pi}{\sqrt{\epsilon_0}}\right)}{2f_2(\tilde{w},\epsilon_0) \frac{\partial}{\partial \tilde w} f_2(\tilde w,\epsilon_0) }\;\!.
\end{align}
In Eq.~(\ref{eq:eps1}) we can substitute $\epsilon_1=\epsilon_0+\delta$, where $\delta$ is small and expand the obtained expression to the linear order in $\delta$. Accounting for Eq.~(\ref{eq:eps0wronskian}) we can solve the obtained expression and obtain $\delta$ that is given by
\begin{gather}\label{eq:deltageneralold}
\epsilon_1-\epsilon_0= \frac{\sqrt{\epsilon_1}\;\!\exp\left(-\frac{\pi}{\sqrt{\epsilon_1}}\right)}{A(\epsilon_1) \frac{\partial}{\partial \tilde w} [f_2(\tilde w,\epsilon_1)]^2 },
\end{gather}
where
\begin{gather}
A(\epsilon)=\frac{\partial}{\partial\epsilon}\left[\frac{f_1(\tilde w,\epsilon)}{f_2(\tilde w,\epsilon)}- \frac{f_1(r_s+\tilde w,\epsilon)}{f_2(r_s+\tilde w,\epsilon)}\right].
\end{gather}
Note that in the right-hand side of Eq.~(\ref{eq:deltageneralold}) we eventually replaced $\epsilon_0$ by $\epsilon_1$, which is possible at $r_s\gg 1$ since in this case the difference between $\epsilon_0$ and $\epsilon_1$ is exponentially small.

Let us find $\epsilon_1$ for values of $\tilde w$ that are not particularly small. In this case we can use the result (\ref{eq:erdelyi}) in both sides of Eq.~(\ref{eq:eps1}). It yields
\begin{align}\label{eq:eps1wlarge}
\frac{\Ai\biglb(g(\epsilon_1,r_s+\tilde w) \bigrb)}{\Bi\biglb(g(\epsilon_1,r_s+\tilde w) \bigrb)} = \frac{\Ai\biglb(g(\epsilon_1,\tilde w) \bigrb)}{\Bi\biglb(g(\epsilon_1,\tilde w) \bigrb)},
\end{align}
where the notation of Eq.~(\ref{eq:g}) is employed. At the leading order in large $r_s$, we have $\epsilon_1=4/(r_s+\tilde w)$ such that the argument of $\phi$-function inside $g(\epsilon_1,r_s+\tilde w)$ in the left-hand side of Eq.~(\ref{eq:eps1wlarge}) equals one. For any $\tilde w$ the argument of $\phi$-function inside $g(\epsilon_1,\tilde w)$ is $\tilde w/(r_s+\tilde w)<1$. Since $\phi(x)<0$ at $x<1$, see Eq.~(\ref{eq:phi<}), we can conclude that $g(\epsilon_1,\tilde w)$ is positive and scales as $r_s^{1/3}$ at a fixed $w$. Therefore, the argument of $\Ai$- and $\Bi$-functions on the right-hand side of Eq.~(\ref{eq:eps1wlarge}) is positive and large. Using the asymptotic expansions for $x\gg 1$,
\begin{gather}
\Ai(x)=\frac{e^{-\frac{2}{3}x^{3/2}}}{2\sqrt\pi x^{1/4}}\left[1+\mathcal{O}(x^{-3/2}) \right],\\
\Bi(x)=\frac{e^{\frac{2}{3}x^{3/2}}}{\sqrt\pi x^{1/4}}\left[1+\mathcal{O}(x^{-3/2}) \right],
\end{gather}
we obtain that the right-hand side of Eq.~(\ref{eq:eps1wlarge}) is exponentially small. Equation~(\ref{eq:eps1wlarge}) thus simplifies to
\begin{align}\label{eq:aizerogeneral}
\Ai\biglb(g(\epsilon_1,r_s+\tilde w) \bigrb)=0.
\end{align} 
As we already discussed, the lowest-energy solution $\epsilon_1$ satisfies
\begin{align}
g(\epsilon_1,r_s+\tilde w)=a_1,
\end{align}
where $a_1$ is the first zero of the Airy-$\Ai$ function. The solution is
\begin{align}\label{eq:eps1www}
\epsilon_1=\frac{4}{r_s+\tilde w}-\frac{4a_1}{(r_s+\tilde w)^{4/3}}+\mathcal{O}\biglb((r_s+\tilde w)^{-5/3}\bigrb),
\end{align}
which reduces to Eq.~(\ref{eq:eps1solution}) at small $\tilde w$.

The evaluation of Eq.~(\ref{eq:deltageneralold}) greatly simplifies once the condition (\ref{eq:aizerogeneral}) as well of the smallness of the right-hand side of Eq.~(\ref{eq:eps1wlarge}) are accounted for. In particular we find
\begin{align}
\frac{\partial}{\partial \tilde w} [f_2(\tilde w,\epsilon)]^2={}& - \sqrt{\epsilon}\;\!\exp\Bigglb({-\frac{\pi}{\sqrt\epsilon}+\frac{8 \left[-\phi\left(\frac{\epsilon\;\!\tilde w}{4}\right)\right]^{3/2}}{3\sqrt\epsilon}}\Biggrb)\notag\\
&\times\cos^2 \biglb(\alpha(\epsilon)\bigrb)
\end{align}
within the leading pre-exponential term and
\begin{align}
A(\epsilon)=\frac{\Ai'(a_1)}{-\Bi(a_1)}\frac{\frac{\partial}{\partial\epsilon} g(\epsilon,r_s+\tilde{w})}{\cos^2 \biglb(\alpha(\epsilon)\bigrb)}.
\end{align}
Using $\frac{\partial}{\partial\epsilon} g(\epsilon,r_s+\tilde{w})|_{\epsilon=\epsilon_1}=2^{2/3}/\epsilon_1^{4/3}$ we eventually obtain
\begin{align}\label{eq:e1e0}
\epsilon_1-\epsilon_0=\frac{-\Bi(a_1)}{2^{2/3}\Ai'(a_1)} \epsilon_1^{4/3} \exp\Bigglb(-\frac{8 \left[-\phi\left(\frac{\epsilon_1\tilde w}{4}\right)\right]^{3/2}}{3\sqrt{\epsilon_1}} \Biggrb),
\end{align}
where $\epsilon_1$ should be taken from Eq.~(\ref{eq:eps1www}) and the $\phi$-function is given by Eq.~(\ref{eq:phi<}). Equation (\ref{eq:e1e0}) accounts for the leading pre-exponential factor in small $\epsilon_1$. In the exponent, however we can have two leading orders. In the regime $r_s\gg\tilde{w}\gg 1$, the final expression is given by
\begin{align}\label{eq:e1e0final}
	\epsilon_1-\epsilon_0=\frac{-4\Bi(a_1)}{\Ai'(a_1)} \frac{ \exp\left( -\pi\sqrt{r_s}-\frac{\pi a_1}{2} r_s^{1/6}+4\sqrt{\tilde{w}} \right)}{r_s^{4/3}}.
\end{align} 
In comparison to the previous result (\ref{eq:Deltafinal}), the dependence on $r_s$ is the same. On the other hand, the dependence on the width $\tilde w$ is rather different. While the splitting (\ref{eq:Deltafinal}) vanishes as $\tilde w\to 0$, the result (\ref{eq:e1e0final}) only applies at sufficiently large $\tilde{w}$. The latter shows that the splitting shows a subexponential growth of $\tilde w$. The increase of the splitting with increasing $\tilde w$ is physically expected as in this case the two wells become less isolated that results with a higher overlap of the symmetric and antisymmetric states.

\section{WKB solution}\label{sec6}

\begin{figure}
\includegraphics[width=0.95\columnwidth]{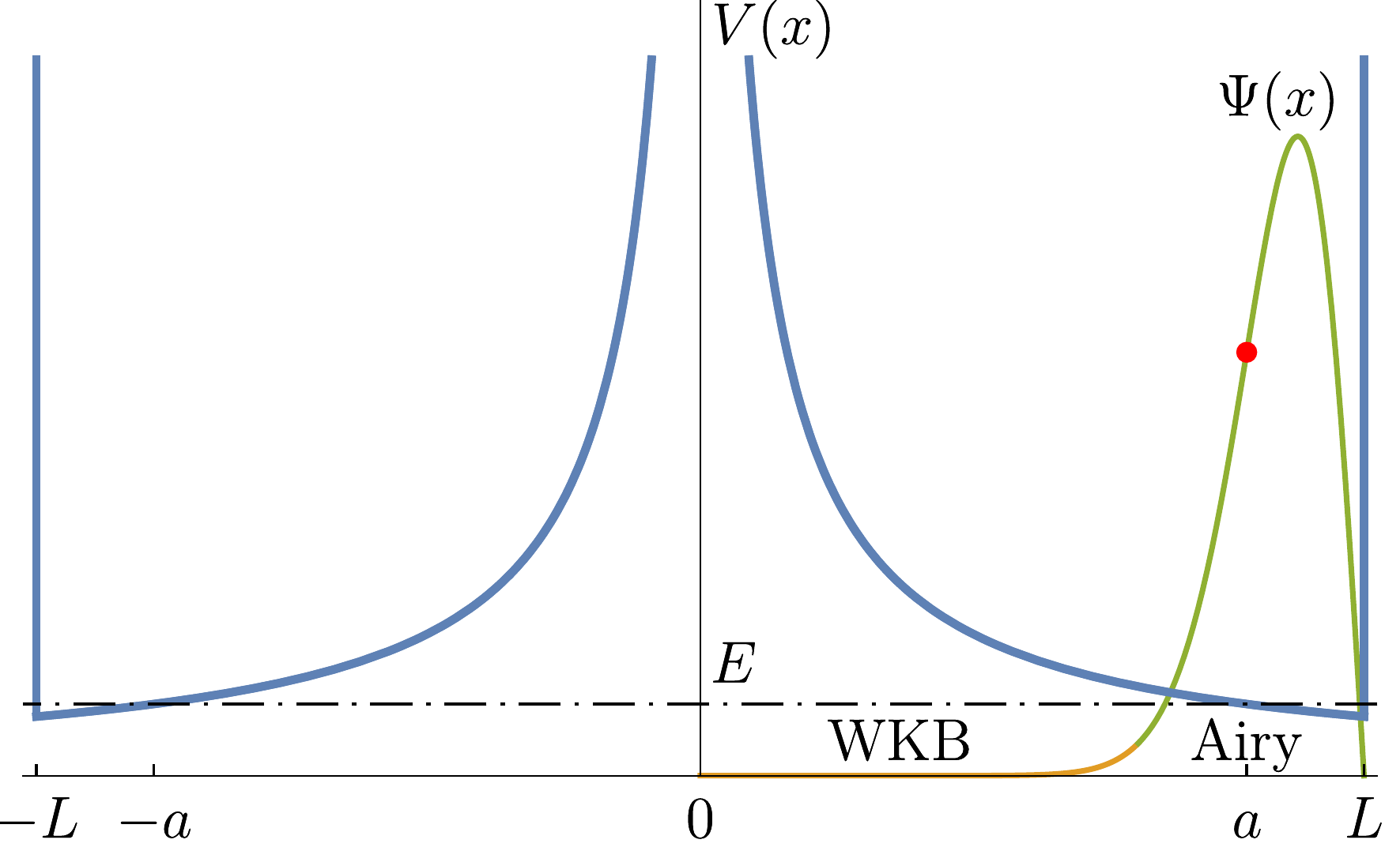}
\caption{Schematic plot of the double well potential $V(x)$ and the wave function $\Psi(x)$ corresponding to the state localized in the right-hand potential well. It has two distinct regions. The one around the turning point $a$, determined by the energy condition $E=V(a)$, is well described by the Airy function, which has a characteristic size $\xi\propto  L^{2/3}$. The other region is left from the turning point at distances beyond $\xi$, where the Airy wave function becomes very small. There the WKB wave function describes well the true wave function of the system.}\label{fig2}
\end{figure}

The problem of finding the difference of the two lowest-energy states for the problem (\ref{eq:SEeffective}) with the potential (\ref{eq:Coulombinteraction}) can also be treated using the quasiclassical approximation.  Our starting point is the expression \cite{Landau3,garg_tunnel_2000}
\begin{align}\label{eq:deltaWKB}
\Delta_{\mathrm{\wkb}}=\frac{4\hbar^2}{m}\Psi(0)\Psi'(0)
\end{align}
for the energy splitting, where $\Psi(x)$ is an approximate ground-state solution of the Schr\"{o}dinger equation with the energy $E$ in the right-hand well of the potential $V(x)$, see Fig.~\ref{fig2}. The quasiclassical wave function $\Psi(x)$ is assumed to be normalized so that the integral of $|\Psi(x)|^2$ over the right well is unity. Moreover, $\Psi(x)$  vanishes at the right edge $x=L$ due to an infinitely strong potential at $x>L$ and decays exponentially on the other side.

The quasiclassical  solution of the eigenvalue problem (\ref{eq:SEeffective}) in the right-hand well can be constructed by a combination of the WKB wave function from one side and the Airy wave function from the other. The present problem has a hard wall boundary condition that imposes zero probability of finding the particle outside the compact interval. Note that this situation is quite different from the more common case of the ``soft'' double well potential where the probability to find the particle is nonzero everywhere in space \cite{garg_tunnel_2000}. Assume the energy level $E$ crosses the potential curve of the right well at $x=a$, see Fig.~\ref{fig2}. This provides a definition for $a$, which is given by the condition
\begin{align}\label{eq:level}
	E=V(a).
\end{align} 
Introducing $x=a+h$, we can linearize the potential around $a$, and study the  wave function in the neighborhood of $a$ in terms of $\omega(h)\equiv \Psi(a+h)$. The latter obeys the Airy differential equation
\begin{align}\label{eq:Airy}
\frac{d^2\omega(h)}{dh^2}+\frac{h}{\xi^3}\omega(h)=0.
\end{align} 
Here we have introduced the length scale $\xi$ by
\begin{align}
\xi=\left(-\frac{\hbar^2}{mV'(a)}\right)^{1/3},
\end{align}	
which determines the characteristic size where $\Psi(x)$ is appreciable. Physically acceptable solution  of Eq.~(\ref{eq:Airy}) is given by
\begin{align}\label{eq:Airywf}
\omega(h)=c \Ai\left(-h/\xi\right),
\end{align}
where $c$ is the normalization factor. Accounting for the boundary condition $\Psi(L)=0$ we find that the inflection point $a$ of the wave function is given by
\begin{align}\label{eq:a1}
	a=L+a_1\xi.
\end{align} 
For the Coulomb potential (\ref{eq:Coulombinteraction}), we then obtain
\begin{align}\label{eq:xi}
\xi=a_{_B} \left(\frac{a+w}{a_{\B}}\right)^{2/3}\simeq a_{\B} \left(\frac{L}{a_{\B}}\right)^{2/3},
\end{align}
where the last equality assumes $L\gg w$. Equation (\ref{eq:xi}) shows that the characteristic length scale  of the wave function scales with the system size as $L^{2/3}$. For long systems, $L\gg a_{\B}$, it follows $\xi\ll L$ and therefore $\Psi(x)$ is localized near the edge. For the energy level (\ref{eq:level}) we find
\begin{align}
E=V(L+a_1\xi)\simeq \frac{e^2}{L}-a_1\frac{e^2 a_{\B}^{1/3}}{L^{4/3}},
\end{align} 
which is in agreement with the result (\ref{eq:eps1solution}) after restoring the notation of Eq.~(\ref{eq:units}). 

At distances $-h\gg\xi$, denoting the region far to the left from the inflection point, the wave function (\ref{eq:Airywf}) decays exponentially,
\begin{align}\label{eq:Airyfar}
\omega(h)=\frac{c}{2\sqrt\pi} \left(-\frac{\xi}{h}\right)^{1/4} e^{-\frac{2}{3}\left(-\frac{h}{\xi}\right)^{3/2}}.
\end{align}
Therefore, the wave function is very small outside its characteristic region of the size $\xi$ around the turning point. This enables us to calculate the normalization factor by integrating the wave function over the whole right-hand well. This yields
\begin{align}
c=\frac{1}{\sqrt\xi \Ai'(a_1)}.
\end{align} 
Here the value of the derivative of the Airy function at its first zero is $\Ai'(a_1)\approx 0.7012$.

In the region $x<a$, the wave function can be approximated by the standard WKB (or quasiclassical) expression \cite{Landau3}
\begin{align}\label{eq:wkb}
\Psi_{\mathrm{\wkb}}(x)=C\frac{\sqrt\hbar}{\sqrt{p(x)}} \exp\Bigglb({\frac{1}{\hbar}\int_0^{x} dy\;\! p(y)}\Biggrb).
\end{align}
Here $C$ is a constant and $p(x)=\sqrt{m[V(x)-E]}$ with $E=V(a)$. Equation~(\ref{eq:wkb}) is a good approximation of the true wave function provided the accuracy condition for the WKB approximation,
\begin{align}\label{eq:accuracycondition}
\hbar \left|\frac{dp(x)}{dx}\right|\ll p^2,
\end{align}
 is satisfied. For $x$ around $a$ it reduces to $a-x\gg \xi$, which denotes that $x$ that is around $a$ should nevertheless be sufficiently far from $a$. On the other hand, Eq.~(\ref{eq:accuracycondition}) near the origin imposes $w\gg a_{\B}$. This denotes that the potential $V(x)$ should not be too steep near the origin in order that the quasiclassical approximation becomes applicable. 

The argument in the exponential function of Eq.~(\ref{eq:wkb}) can be decomposed as
\begin{align}
{\frac{1}{\hbar}\int_0^{x} dy\;\! p(y)}={\frac{1}{\hbar}\int_0^{a} dy\;\! p(y)}-{\frac{1}{\hbar}\int_{x}^{a} dy\;\! p(y)}.
\end{align}
For $x$ in the vicinity of $a$, the latter integral can be evaluated after the linearization of the integrand. There we use $p(x)/\hbar\simeq \sqrt{a-x}/\xi^{3/2}$ to find
\begin{align}\label{eq:exp}
{\frac{1}{\hbar}\int_{x}^{a} dy\;\! p(y)}\simeq\frac{2}{3}\left(\frac{a-x}{\xi}\right)^{3/2}. 
\end{align}
Equation (\ref{eq:exp}) coincides with the argument of the exponential in Eq.~(\ref{eq:Airyfar}), whereas $\sqrt{p(x)}$ of Eq.~(\ref{eq:wkb}) agrees with the prefactor in front of the exponential. For 
\begin{align}
C=\frac{c}{2\sqrt{\pi\xi}} \exp\Bigglb(-\frac{1}{\hbar}\int_0^a dy\;\! p(y)\Biggrb)
\end{align}
we thus obtain the matching between the Airy wave function (\ref{eq:Airywf}) that describes well the true wave function  around $a$ in the characteristic region of the size $\xi$, and the WKB wave function (\ref{eq:wkb}) that applies at $a-x\gg\xi$.

Within the accuracy of the WKB approximation we have $\Psi_{\mathrm{\wkb}}'(x)\simeq \Psi_{\mathrm{\wkb}}(x) p(x)/\hbar$. This yields the final expression for the energy splitting (\ref{eq:deltaWKB}) in the form
\begin{align}\label{eq:Deltawkbfinal}
\Delta_{\mathrm{WKB}}=\frac{\hbar^2}{\pi [\Ai'(a_1)]^2 m\xi^2} \exp\Bigglb(-\frac{1}{\hbar}\int_{-a}^{a} dy \;\!p(y)\Biggrb).
\end{align}
For the potential (\ref{eq:Coulombinteraction}), the integral in Eq.~(\ref{eq:Deltawkbfinal})  is elementary,
\begin{align}\label{eq:Deltawkb2}
\frac{1}{\hbar}\int_{-a}^{a} dy \;\!p(y)=2\sqrt{\frac{a+w}{a_{\B}}} \arctan\sqrt{\frac{a}{w}}
-2\sqrt{\frac{aw}{a_{\B}(a+w)}}. 
\end{align}
At $L\gg w\gg a_{\B}$, where the latter is the accuracy condition of the WKB wave function, keeping the terms that do not tend to zero in the exponential function, we obtain
\begin{align}\label{eq:DeltaWKBfinalevaluated}
\Delta_{\mathrm{WKB}}={}&\frac{\hbar^2}{\pi [\Ai'(a_1)]^2 m a_{\B}^2}\notag\\
&\times \frac{\exp\left(-\pi\sqrt{r_s}-\frac{\pi a_1}{2}r_s^{1/6}+4\sqrt{\tilde{w}}\,\right)}{r_s^{4/3}}.
\end{align}
Here the notation of Eq.~(\ref{eq:units}) is used. Equation (\ref{eq:DeltaWKBfinalevaluated}) is the result for the energy splitting in the WKB approximation. It applies for $\tilde{w}\gg 1$, which is complementary to Eq.~(\ref{eq:Deltafinal}). On the other hand, the previously derived expression (\ref{eq:e1e0final}) applies in the same parameter regime as Eq.~(\ref{eq:DeltaWKBfinalevaluated}). It would thus be natural to compare the two. Accounting for the units of Eq.~(\ref{eq:units}) the corresponding ratio of the two energy splittings is
\begin{align}
\left(\frac{-4\Bi(a_1)}{\Ai'(a_1)}\right)\bigg{/}\left(\frac{4}{\pi [\Ai'(a_1)]^2 }\right)= -\pi \Bi(a_1)\Ai'(a_1).
\end{align}
The latter constant in fact is equal to $1$, as it follows, e.g., from the formula for the Wronskian of $\Ai(x)$ and $\Bi(x)$ that is $1/\pi$. Setting $x=a_1$ in the Wronskian, the wanted relation follows. We have therefore shown that Eqs.~(\ref{eq:e1e0final}) and (\ref{eq:DeltaWKBfinalevaluated}) are equivalent.

\section{Discussions}\label{sec7}

\begin{figure}
	\includegraphics[width=0.95\columnwidth]{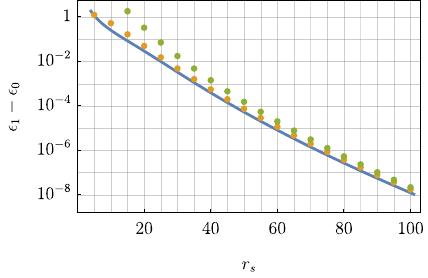}
	\caption{Plot of the energy gap as a function of $r_s$ for $\tilde{w}=10$. The curve represents the exact value obtained by numerically solving Eqs.~(\ref{eq:eps0}) and (\ref{eq:eps1}). The dots that are closer to the curve are obtained from the analytical expression (\ref{eq:deltageneral}). The dots that are less close to the curve are obtained from Eq.~(\ref{eq:e1e0final}) multiplied by $\exp\Bigl(a_1\sqrt{w}/r_s^{1/3}\Bigr)$. The latter corresponds to the subleading contribution involving the width of the wire, which is contained in Eq.~(\ref{eq:deltageneral}).}\label{fig3}
\end{figure}

Previously we only found explicit results for the energy splitting in the case $r_s\gg \tilde{w}$, see Eqs.~(\ref{eq:Deltafinal}) and (\ref{eq:e1e0final}). The derived expression (\ref{eq:e1e0}) is more general and enables us to find
\begin{widetext}
\begin{align}\label{eq:deltageneral}
\epsilon_1-\epsilon_0=\frac{4}{\pi[\Ai'(a_1)]^2} \frac{ \exp\bigglb( -2\sqrt{r_s+\tilde{w}}\left[1+\frac{a_1}{2 (r_s+\tilde{w})^{1/3}}\right]\arctan\sqrt{\frac{r_s}{\tilde w}} +2 \left[1-\frac{a_1}{2 (r_s+\tilde{w})^{1/3}}\right]\sqrt{\frac{r_s\tilde w}{r_s+\tilde w}}\,\biggrb)}{(r_s+\tilde w)^{4/3}},
\end{align}
\end{widetext}
which applies at any ratio $r_s/\tilde{w}$. In the special case $r_s/\tilde{w}\gg 1$, the result (\ref{eq:deltageneral}) reduces to Eq.~(\ref{eq:e1e0final}). In Fig.~\ref{fig3} we show the energy gap for $\tilde w=10$.

One may wonder what happens with the WKB result (\ref{eq:Deltawkbfinal}) for the splitting at arbitrary ratio $r_s/\tilde w$. To this end we need the expression
\begin{align}
a=a_B\left[r_s+a_1 \left(r_s+\tilde w\right)^{2/3}\right],
\end{align}
which follows from Eqs.~(\ref{eq:a1}) and (\ref{eq:xi}). Substituting it into Eq.~(\ref{eq:Deltawkbfinal}) and expanding the obtained expression to the linear order with respect to the term controlled by $a_1$, i.e., the subleading one, we obtain the result that coincides with Eq.~(\ref{eq:deltageneral}). This is a remarkable agreement as the origins of the two expressions are quite different. One is based on the exact solution of the Schr\"{o}dinger equation in terms of the Whittaker functions and the analysis by making use of their limiting form found by Erd\'{e}lyi and Swanson of Eq.~(\ref{eq:erdelyi}); the other is based on the WKB approximation in conjunction with the formula (\ref{eq:deltaWKB}). 

Equation (\ref{eq:deltageneral}) has another interesting limiting case. For $\tilde w\gg r_s\gg 1$ it reduces to
\begin{align}\label{eq:triangular}
\epsilon_1-\epsilon_0=\frac{4}{\pi[\Ai'(a_1)]^2} \frac{ \exp\left(-\frac{4r_s^{3/2}}{3\tilde w}-2a_1 \frac{\sqrt{r_s}}{\tilde{w}^{1/3}}\right)}{\tilde w^{4/3}}.
\end{align}
Equation (\ref{eq:triangular}) supplemented by the condition $r_s\gg w^{2/3}$ describes the energy splitting between the two lowest-energy stats in a triangular double well potential described by
\begin{align}
V(x)=\begin{cases}
-\frac{e^2}{{w}^2}|x|, & -L\le x\le L\\
+\infty, & |x|>L
\end{cases}.
\end{align}
The latter condition arises from  $L\gg \xi$, see Eq.~(\ref{eq:xi}) and Fig.~\ref{fig2}, which is a necessary condition that the triangular well has a well-localized (bound) state in each of the two isolated triangular wells. A physical argument to support this is as follows. Let us consider a particle in, say, the right-hand well and require the particle be localized. This happens if its energy that is on the order of $\hbar^2/mL^2$ is smaller than the well depth, $e^2 L/w^2$. This yields $a_{\B}w^2\ll L^3$, which is equivalent to the above condition.

The eigenenergy of the first excited state was calculated in Eq.~(\ref{eq:eps1solution}). It has direct relation with the first zero of the Airy-$\Ai$ function $a_1$. It is worth mentioning an alternative way to treat  Eq.~(\ref{eq:eps1}) and obtain $\epsilon_1$. 
Consider for simplicity $\tilde{w}\ll 1$ case for which it is sufficient to set $\tilde{w}=0$ for odd eigenfunctions. They have nodes at $x=0$ and are thus not significantly affected by small $\tilde w$. The left-hand side of Eq.~(\ref{eq:eps1}) we represent by Eq.~(\ref{eq:ratioerdelyi}) and the right-hand side by the expression (\ref{eq:f1of2}). Since we study the case of strong repulsion, $r_s\gg 1$, the lowest eigenenergies are small. In this case we can assume that $g(\epsilon_1,r_s)$ of Eq.~(\ref{eq:ratioerdelyi}) is large and  negative. Approximating the ratio of the Airy functions as
\begin{align}
\frac{\Ai(-z)}{\Bi(-z)}\simeq \tan\left(\frac{2z^{3/2}}{3}+\frac{\pi}{4}\right),\quad z\to+\infty,
\end{align}
we obtain
\begin{align}\label{eq:lhs}
\frac{f_1(r_s,\epsilon_1)}{f_2(r_s,\epsilon_1)}\simeq{} \tan\bigglb(\frac{4 \phi^{3/2}\left(\frac{r_s \epsilon_1}{4}\right)}{3\sqrt{\epsilon_1}}-\alpha(\epsilon_1)+\frac{\pi}{4}\biggrb).
\end{align}
Equating the two tangent functions, after the linearization of $\phi$-function according to (\ref{eq:phiexp}), for the lowest energy $\epsilon_1$ we obtain the equation
\begin{align}
	\left(\frac{r_s\;\! \epsilon_1}{4}-1\right)^{3/2}=\frac{9\pi}{16}\sqrt{\epsilon_1}.
\end{align}
Its solution has the form of Eq.~(\ref{eq:eps1solution}) where $-a_1$ should be replaced by $(9\pi/8)^{2/3}$. Comparing the numerical values, $-a_1\approx 2.338$ and $(9\pi/8)^{2/3}\approx 2.320$ we observe a relatively small difference of the numerical coefficient in the subleading term. Our initial assumption that $g(\epsilon_1,r_s)$ is large and negative is not entirely correct as we can \emph{a posteriori} obtain $g(\epsilon_1,r_s)=a_1$ at $r_s\gg 1$, which in fact is not very large. Nevertheless, the obtained, somewhat uncontrolled, result for $\epsilon_1$ is approximately correct. Performing the numerical analysis of Eq.~(\ref{eq:eps1}) we have confirmed that the two leading asymptotic terms given by Eq.~(\ref{eq:eps1solution}) are exact at $r_s\to\infty$.

The present study can be directly generalized to study the splitting of higher energy states. For example, in the case $\tilde w\ll 1$ we can replace $a_1$ by $a_2\approx -4.088$ in Eq.~(\ref{eq:eps1solution}) and obtain the energy of the second lowest odd eigenvalue $\epsilon_3$. Here $a_2$ is the second zero of the Airy-$\Ai$ function. Equation (\ref{eq:Deltafinal}) with the same replacement should then correspond to the difference $\epsilon_3-\epsilon_2$, etc.  

The approach developed in this work can be applied to calculate the energy splitting in hard-wall double-well potentials different than the Coulomb case (\ref{eq:Coulombinteraction}). We note that Eqs.~(\ref{eq:eps0}) and (\ref{eq:eps1}) do not rely on specific form of the two independent solutions of the Schr\"{o}dinger equation. Their Wronskian can be used to connect the right-hand sides of  Eqs.~(\ref{eq:eps0}) and (\ref{eq:eps1}) and express the splitting following the considerations of Sec.~\ref{sec5}. Similarly, the approach of Sec.~\ref{sec6} is also robust to various other forms of the potential. An example where the solution appears to be straightforward is the inverse square potential $V(x)\sim (w+|x|)^{-2}$. Two independent solutions of the corresponding eigenvalue problem are given in terms of the Bessel functions of the first and second kind. We defer this problem for future studies. 

This study leaves open some interesting questions. In longer Wigner crystals the leading exponential term in the exchange coupling is controlled by the coefficient $\eta\approx 2.798$, see Sec.~\ref{sec1}, while in the present study we have found $\eta=\pi$ for a system with two electrons and zero boundary conditions. Increasing the number of electrons in the present setup, it is expected that the boundary conditions will stop playing a role and therefore the coefficient $\eta=\pi$ should drift toward $\eta\approx 2.798$ for a larger number of electrons in the middle of the system. Namely the exchanging electrons in this case will have neighbors that will have the role of ``soft'' walls therefore increasing effectively the available space for the exchanging electrons. On the other hand the role of the subleading exponential term controlled by $\tilde \eta$ is also expected to diminish since its scaling $r_s^{1/6}$ arises from the properties of the Airy functions that should not be relevant in the middle of the system if it has a sufficient number of electrons. Another open questions is understanding the role of quantum fluctuations of the center of mass coordinate in the present setup. In other setups, its role was marginal resulting with the decrease of the exponent less than $1\%$ and the increase of the overall prefactor by $50\%$~\cite{fogler_exchange_2005}.

In summary, we obtained analytical results for the exchange coupling of two electrons in a one-dimensional system with zero boundary conditions, which resembles the realistic system realized experimentally. We showed that in addition to the leading exponentially decaying contribution as a function of the distance, there is a subleading term with exponential scaling that significantly increases the exchange coupling, thus playing an important role.

\section*{Acknowledgments}

This project has received financial support from the
CNRS through the MITI interdisciplinary programs.


%

\end{document}